\documentclass{article}
\usepackage{spconf,algorithm,algorithmic,amsmath,amssymb,amsthm,bbm,cite,color,graphicx,microtype,url}
\usepackage[font=small]{caption}

\DeclareSymbolFont{matha}{OML}{txmi}{m}{it}
\DeclareMathSymbol{\varv}{\mathord}{matha}{118}
\newtheorem{prop}{Proposition}

\newcommand{\Exp}{\mathbb{E}}
\newcommand{\herm}{\mathrm{H}}
\renewcommand{\Im}{\mathrm{Im}}
\renewcommand{\Re}{\mathrm{Re}}
\newcommand{\sgn}{\mathrm{sgn}}

\newcommand{\Diag}{\mathrm{Diag}}
\newcommand{\st}{\mathrm{s.t.}}
\newcommand{\tr}{\mathrm{Tr}}
\newcommand{\tran}{\mathrm{T}}

\title{Metasurface-Based Receivers with 1-bit {ADCs}\\ for Multi-User Uplink Communications}

\name{Panagiotis~Gavriilidis,$^1$ Italo~Atzeni,$^2$ and George~C.~Alexandropoulos$^1$
\thanks{The work of P.~Gavriilidis and G.~C.~Alexandropoulos was supported by the Smart Networks and Services Joint Undertaking (SNS JU) project TERRAMETA under the European Union's Horizon Europe research and innovation programme under Grant Agreement no. 101097101, including top-up funding by UK Research and Innovation (UKRI) under the UK government's Horizon Europe funding guarantee. The work of I.~Atzeni was supported by the Research Council of Finland (336449 Profi6, 346208 6G~Flagship, 348396 HIGH-6G, and 357504 EETCAMD).}}

\address{$^1$Dept. of Informatics and Telecommunications, National and Kapodistrian University of Athens, Greece \\
$^2$Centre for Wireless Communications, University of Oulu, Finland}

\begin{document}
%
\maketitle

\begin{abstract}
\vspace{-1mm}
The massive Multiple-Input Multiple-Output (mMIMO) concept has been recently moving forward to extreme scales to address the envisioned requirements of next generation networks. However, the extension of conventional architectures will result in significant cost and power consumption. To this end, metasurface-based transceivers, consisting of microstrips of metamaterials, have recently emerged as an efficient enabler of extreme mMIMO systems. In this paper, we consider metasurface-based receivers with a 1-bit Analog-to-Digital Converter (ADC) per microstrip and develop an analytical framework for the optimization of the analog and digital combining matrices. Our numerical results, including comparisons with fully digital, infinite-resolution MIMO, provide useful insights into the role of various system parameters.
\end{abstract}

\vspace{-1mm}
\section{Introduction}
\label{sec:intro}
\vspace{-2mm}

Reconfigurable intelligent surfaces have arisen as one of the most prominent technologies for future 6th-generation networks due to their appealing characteristics \cite{CE_overview_2022,HMIMO_survey_all}. They are dynamically reconfigurable, a feature that allows for a favorable design of the propagation environment under various scenarios \cite{marconi_george,Strinati2021Reconfigurable,Space_shift_keying_RIS,RIS_pervasive}. Furthermore, they interact with Electro-Magnetic (EM) waves in the respective domain~\cite{PhysFad}, eliminating the need for transmit or Receive (RX) Radio-Frequency (RF) chains along with the corresponding signal amplifiers and converters, thus, substantially decreasing the power consumption and cost \cite{HMIMO_George}.
Aiming to design a low-cost metasurface-based massive multiple-input multiple-output (mMIMO) transceiver, \cite{DMA_origin} introduced Dynamic Metasurface Antennas (DMAs) and studied efficient beamforming designs. Further research on DMAs under various scenarios, such as THz communications \cite{gavras2023duplex}, radio environment sensing \cite{DMA_sensing}, and low-resolution Analog-to-Digital Converters (ADCs) \cite{DMA_quant_optimization,gavras2024dma}, has recently highlighted the potential of this family of transceiver architectures.

Numerous recent works have considered low-resolution ADCs to reduce the power consumption of the RX RF chains in mMIMO architectures \cite{throughput_analysis_low_bit,Channel_est_1_bit,Atz22,DMA_quant_optimization,THz_1_bit}, since the ADCs' power consumption scales linearly with the bandwidth and exponentially with the number of resolution bits \cite{ADC_FOM_Power_consumpt,Atz21}. The work in \cite{throughput_analysis_low_bit} was among the first to provide a throughput analysis for low-resolution ADCs, investigating the conditions under which they can approach the achievable rate of the infinite-resolution case.
Channel estimation and data detection in mMIMO with 1-bit ADCs were investigated in \cite{Channel_est_1_bit,Atz22}. 1-bit THz mMIMO systems were studied in \cite{THz_1_bit}, which proposed an optimization scheme for transmit beamforming using unsupervised learning. DMAs with orthogonal frequency division multiplexing and low-resolution ADCs were considered in \cite{DMA_quant_optimization}, which presented a task-based quantization scheme for symbol detection.

In this paper, we study RX DMAs with 1-bit ADCs for multi-user uplink communications. Differently from \cite{DMA_quant_optimization}, we consider element-wise quantization at each RX RF chain and rate optimization. Capitalizing on the Bussgang decomposition, we present an analytical framework for assessing the achievable sum-rate performance. The derived closed-form expression for this metric is then used to obtain the optimal analog and digital combiners. Our numerical investigation, including comparisons with fully digital, infinite-resolution MIMO, showcases the trade-offs among the number of RX RF chains and the number of elements per microstrip.

\vspace{-1mm}
\section{System model}
\label{sec:system_model}
\vspace{-2mm}

Consider a DMA-based Base Station (BS) wishing to serve \(K\) single-antenna users in the uplink. The RX DMA at the BS includes a total of \(N\triangleq N_d N_e\) metamaterial elements, grouped into \(N_d\) microstrips with \(N_e\) elements each. The channel between the BS and the users is denoted by \(\mathbf{H}\triangleq [\mathbf{h}_1,\ldots,\mathbf{h}_K]\in \mathbb{C}^{N \times K}\), where \(\mathbf{h}_i\) represents the channel between the BS and the $i$th ($i=1,\ldots,K$) user. Moreover, let \(\mathbf{x}\triangleq [x_1,\dots,x_K]^T\in \mathbb{C}^{K \times 1}\) and \(\mathbf{n}\in \mathbb{C}^{N \times 1}\) denote the transmit symbol and the thermal noise vectors, respectively. We assume that the elements of $\mathbf{H}$, $\mathbf{x}$, and $\mathbf{n}$ are independent standard complex normal Random Variables (RVs).
Furthermore, we set a common transmit signal-to-noise ratio (SNR) \(\rho\), which incorporates large-scale phenomena for all the users (as in \cite{throughput_analysis_low_bit,Channel_est_1_bit,Atz22}).\footnote{This assumption is valid for the 1-bit quantization considered in this paper, since there is no signal dynamic range and, thus, a power control scheme needs to be applied so that the strong users do not suppress the weaker ones \cite{Channel_est_1_bit}.} The impinging signal at the RX DMA can be mathematically expressed in baseband as follows:
\vspace{-1mm}
\begin{equation} \label{eq:DMA_surf_signal}
    \mathbf{y} \triangleq \sqrt{\rho}\mathbf{H}\mathbf{x}+\mathbf{n} \in \mathbb{C}^{N \times 1}.
\end{equation}
The $n$th ($n=1,\ldots,N$) element of $\mathbf{y}$ is multiplied by an analog weight and propagates within the respective microstrip. Then, each microstrip outputs a signal that is the superposition of its elements' signals. Let \(\mathbf{Q}\in \mathbb{C}^{N_d \times N}\) model the DMA's tunable analog weights  with \(\mathbf{Q}^H\triangleq{\rm blkdiag}[\mathbf{q}_1,\dots,\mathbf{q}_{N_d}]\). Each \(\mathbf{q}_i\triangleq[q_{i,1},\dots,q_{i,N_e}]^H\) includes the analog weights in each \(i\)th microstrip, which are given by \cite{Xu_DMA_2022}:
\vspace{-1mm}
\begin{equation}\label{eq:Lorentzian_constrained}
    q_{i,l}\in \mathcal{L}\triangleq \left\{\frac{\jmath + e^{\jmath\phi}}{2} \mid \phi \in \left[0,2\pi\right] \right\}.
\end{equation}
We also introduce the diagonal matrix \(\mathbf{A} \in \mathbb{C}^{N \times N}\) accounting for the microstrip propagation, assuming a lossless waveguide, with \(\mathbf{[A]}_{(i-1)N_e+l,(i-1)N_e+l}\triangleq e^{\jmath \beta \ell_{i,l}}\), where \(\beta\) is the microstrip's wavenumber and \(\ell_{i,l}\) denotes the distance of the \(l\)th element in the \(i\)th microstrip to the output port. Putting all the above together, the signal at the output of the $N_d$ microstrips is given in baseband by $\mathbf{z}=\mathbf{QAy}\in \mathbb{C}^{N_d \times 1}$.

In this paper, we consider that the output of each microstrip is fed to an RX RF chain possessing a pair of low-resolution ADCs for its in-phase and quadrature components. To this end, the output of all the ADCs is given by \(\mathbf{r} \triangleq \mathcal{Q} (\mathbf{z}) \in \mathbb{C}^{N_d \times 1}\), where $\mathcal{Q} (\cdot)$ is the quantization function. Following the Gaussian signaling assumption, we can use the Bussgang decomposition \cite{Bussgang_Theory} to express \(\mathbf{r}\) as follows:
\vspace{-1mm}
\begin{equation} \label{eq:r_Bussgang}
\mathbf{r} = \mathbf{G z + d},
\end{equation}
where $\mathbf{d} \in \mathbb{C}^{N \times 1}$ is the zero-mean, non-Gaussian quantization distortion vector that is uncorrelated with $\mathbf{z}$, and $\mathbf{G}$ is the Bussgang gain, i.e., a diagonal matrix given by \(\mathbf{G} \triangleq \mathbf{C}_{\mathbf{zr}}^{\herm} \mathbf{C}_{\mathbf{z}}^{-1}\), with \(\mathbf{C}_{\mathbf{zr}} \triangleq \Exp [\mathbf{z} \mathbf{r}^{\herm}] \) and \( \mathbf{C}_{\mathbf{z}} \triangleq \Exp [\mathbf{zz}^{\herm}] \). Defining \(\mathbf{C}_{\mathbf{r}} \triangleq \Exp [\mathbf{r r}^{\herm}] = \mathbf{G C_{z} G + C_{d}}\) and \(\mathbf{C_{d}} \triangleq \Exp [\mathbf{d d}^{\herm}]= \mathbf{C_{r} - G C_{z} G }\) and using \cite[Theorem~3]{Bussgang_Theory}, the correlation matrix between the received signal \(\mathbf{r}\) and the transmit symbols \(\mathbf{x}\) is given by:
\vspace{-1mm}
\begin{equation}
    \mathbf{C}_{\mathbf{rx}}\!\triangleq\!\Exp [\mathbf{r x}^{\herm}]\!=\!\mathbf{C_{rz}C_{z}}^{-1}\mathbf{C}_{\mathbf{zx}}\!=\!\mathbf{G}\mathbf{C}_{\mathbf{zx}}\!=\!\sqrt{\rho}\mathbf{G}\mathbf{QAH}.
\end{equation}

Building upon the received quantized signal in \eqref{eq:r_Bussgang}, the soft-estimated symbol from the $k$th user can be expressed as:
\vspace{-1mm}
\begin{equation}
\hat{{x}}_{k} = \mathbf{w}_{k}^{\herm} \bigg( \mathbf{GQA}\bigg( \sqrt{\rho} \sum_{\bar{k}=1}^{K} \mathbf{h}_{\bar{k}} x_{\bar{k}} + \mathbf{n} \bigg) + \mathbf{d} \bigg),
\end{equation}
where $\mathbf{W} \triangleq [\mathbf{w}_{1}, \dots, \mathbf{w}_{K}] \in \mathbb{C}^{N_d \times K}$ denotes the digital combining matrix. Hence, the corresponding Signal-to-Interference-plus-Noise-and-Distortion Ratio (SINDR) can be expressed via the following formula:
\vspace{-1mm}
\begin{equation} \label{eq:SINDR_k}
\mathrm{SINDR}_{k} \triangleq \frac{\rho |\mathbf{w}_{k}^{\herm} \mathbf{G Q A h}_{k}|^{2}}{I_k + |\mathbf{w}_{k}^{\herm} \mathbf{G Q A }|^2 + \mathbf{w}_{k}^{\herm} \mathbf{C}_{\mathbf{d}} \mathbf{w}_{k}},
\end{equation}
where $I_k\triangleq\sum_{\bar{k} \neq k} \rho |\mathbf{w}_{k}^{\herm}\mathbf{ G Q A h}_{\bar{k}}|^{2}$ represents the interference term. The achievable sum rate with perfect Channel State Information (CSI) is thus obtained as:
\vspace{-1mm}
\begin{equation}\label{eq:Rate_original}
\mathcal{R} \triangleq \sum_{k=1}^{K} \log_{2} (1 + \mathrm{SINDR}_{k}).
\end{equation}
Since 1-bit ADCs are used at the RX DMA, the quantization function becomes \( \mathcal{Q} (\mathbf{z}) \triangleq \sqrt{\frac{\eta}{2}} \big( \sgn \big( \Re [\mathbf{z}] \big) + \jmath \, \sgn \big( \Im [\mathbf{z}] \big) \big) \) \cite{Atz22}. In this case\footnote{The scaling factor $\eta$ can be chosen such that the variance of the output coincides with that of the input. Note that this does not affect the SINDR.}, $\mathbf{G}$ and $\mathbf{C}_{\mathbf{d}}$ can be computed in closed form, yielding the following expressions: \cite{Channel_est_1_bit}: 
\vspace{-1mm}
\begin{align}
\mathbf{G} & = \sqrt{\frac{2}{\pi} \eta} \Diag (\mathbf{C}_{\mathbf{z}})^{- \frac{1}{2}}, \\
\nonumber \mathbf{C}_{\mathbf{r}} & = \frac{2}{\pi} \eta \Big( \arcsin \big( \Diag (\mathbf{C}_{\mathbf{z}})^{- \frac{1}{2}} \Re [\mathbf{C}_{\mathbf{z}}] \Diag (\mathbf{C}_{\mathbf{z}})^{- \frac{1}{2}} \big) \\
& \phantom{=} \ + \jmath \arcsin \big( \Diag (\mathbf{C}_{\mathbf{z}})^{- \frac{1}{2}} \Im [\mathbf{C}_{\mathbf{z}}] \Diag (\mathbf{C}_{\mathbf{z}})^{- \frac{1}{2}} \big) \Big).
\end{align}
These expressions can be used to compute the SINDR in \eqref{eq:SINDR_k}.

\vspace{-1mm}
\section{Proposed RX DMA Design}
\label{sec:optimization}
\vspace{-2mm}

In this section, we focus on the joint design of the digital combiner \(\mathbf{W}\) and the analog combiner \(\mathbf{Q}\) to maximize the achievable rate in \eqref{eq:Rate_original}. Using the definition \(\mathbf{q}\triangleq[\mathbf{q}_1^{\tran},\dots,\mathbf{q}_{N_d}^{\tran}]^{\tran}\), this problem can be mathematically formulated as:
\vspace{-1mm}
\begin{equation}\label{eq:OP_origin}
\begin{array}{rcl}
\mathcal{OP}: & \displaystyle \max_{\mathbf{W},\mathbf{Q}} & \mathcal{R}(\mathbf{W},\mathbf{Q}) \\
& \st & \mathbf{q}^{\ast}\in \mathcal{L}^{N\times 1}, \, \mathbf{Q}={\rm blkdiag}[\mathbf{q}_1^{\herm},\dots,\mathbf{q}_{N_d}^{\herm}].
\end{array}
\end{equation}
To deal with $\mathcal{OP}$'s non-convexity and structural constraints, we first derive the output SINDR for the optimal \(\mathbf{W}\), and then, we maximize the resulting rate with respect to \(\mathbf{Q}\).

\vspace{-1mm}
\subsection{Design of the Digital Combiner \(\mathbf{W}\)}
\vspace{-1mm}

It can be easily observed that each column \(\mathbf{w}_k\) of the digital combiner can maximize the corresponding SINDR by solving a generalized eigenvalue problem. Specifically, all the SINDRs are of the form \({\rm SINDR}_k=\frac{\mathbf{w}_k^{\herm}\mathbf{N}_k\mathbf{w}}{\mathbf{w}^{\herm}_k\mathbf{D}_k\mathbf{w}_k}\) and, thus, the optimal \(\mathbf{w}_k\) is the principal eigenvector of \(\mathbf{D}_k^{-1/2}\mathbf{N}_k\mathbf{D}_k^{-1/2}\), with $\mathbf{N}_k\triangleq \rho \mathbf{K}\mathbf{h}_k\mathbf{h}_k^{\herm}\mathbf{K}^{\herm}$, $\mathbf{D}_k\triangleq \sum_{\bar{k} \neq k} \rho \mathbf{ K h}_{\bar{k}}\mathbf{h}_{\bar{k}}^{\herm}\mathbf{K}^{\herm} +\mathbf{KK}^{\herm}+\mathbf{C_d}$, and \(\mathbf{K}\triangleq\mathbf{GQA}\), which is scaling invariant. As proved in \cite[Section~6.4.2]{heath2018foundations}, the resulting SINDR is the same as the SINDR at the output of a linear minimum mean squared error filter. Hence, we design the optimal digital combiner $\mathbf{W}$ as:
\vspace{-1mm}
\begin{align}
    \mathbf{W} &\triangleq \mathbf{C}_{\mathbf{r}}^{-1}\mathbf{C}_{\mathbf{rx}}\nonumber\\
    &= \sqrt{\rho} (\rho\mathbf{KH}\mathbf{H}^{\herm}\mathbf{K}^{\herm}+\mathbf{KK}^{\herm}+\mathbf{C}_{\mathbf{d}})^{-1}\mathbf{KH}.
\end{align} 
After some algebraic manipulations following the methodology in \cite[Section~6.4.2]{heath2018foundations}, which are omitted here due to page limitation, the SINDR of each $k$th user is given by:
\vspace{-1mm}
\begin{equation}\label{eq:LMMSE_SINDR}
    {\rm SINDR}_k^{\star}=\mathbf{h}_k^{\herm}\mathbf{K}^{\herm}\bigg(\mathbf{KH}_{\text{-}k}\mathbf{H}_{\text{-}k}^{\herm}\mathbf{K}^{\herm}+\frac{\mathbf{KK}^{\herm}}{\rho}+\frac{\mathbf{C_d}}{\rho}\bigg)^{-1} \mathbf{K}\mathbf{h}_k,
\end{equation}
where \(\mathbf{H}_{\text{-}k}\mathbf{H}_{\text{-}k}^{\herm}=\mathbf{HH}^{\herm}-\mathbf{h}_{k}\mathbf{h}_{k}^{\herm}\).

\vspace{-1mm}
\subsection{Design of the Analog Combiner \(\mathbf{Q}\)}
\vspace{-1mm}

The objective of the analog combiner \(\mathbf{Q}\) is to maximize the achievable rate using the solution for \(\mathbf{W}\) from the previous section.  
We follow the methodologies proposed in \cite{DMA_Optimization,DMA_quant_optimization} which focus on reaching a quadratic form with respect to the stacking vector \(\mathbf{q}\) that includes all the non-zero elements of \(\mathbf{Q}^{\herm}\). To this end, we reformulate the sum-rate objective twice using the auxiliary variables \(\boldsymbol{\gamma} \triangleq [\gamma_{1}, \ldots, \gamma_{K}]^{T} \in \mathbb{R}^{K \times 1}\) and \(\mathbf{Y} \triangleq [\mathbf{y}_{1}, \ldots, \mathbf{y}_{K}] \in \mathbb{C}^{N_d \times K}\). In the first step, by introducing \(\boldsymbol{\gamma}\), we aim to take the ratio out of the \(\log\) function \cite[Proposition~1]{DMA_Optimization}. Subsequently, we disentangle the ratio term by introducing the auxiliary variable \(\mathbf{Y}\), and then, using the matrix quadratic transform \cite[Proposition~2]{DMA_Optimization}, the sum-rate performance can be equivalently expressed as follows:
\vspace{-1mm}
\begin{align}
\nonumber &\mathcal{R}(\mathbf{Q},\!\boldsymbol{\gamma},\!\!\mathbf{Y})\!\!=\!\!\sum_{k=1}^{K}\!\Big(\!\log_2(1\!+\!\gamma_k)\!-\!\gamma_k\!+\!(1\!+\!\gamma_k)\Big(\!2\Re[\mathbf{h}_k^{\herm}\mathbf{K}^{\herm}\mathbf{y}_k]\\
    &- \mathbf{y}_k^{\herm}\big(\mathbf{KHH}^{\herm}\mathbf{K}^{\herm}+\rho^{-1}(\mathbf{KK}^{\herm}+\mathbf{C_d})\big)\mathbf{y}_k\Big)\Big).
\end{align}
Since \(\mathcal{R}(\mathbf{Q},\boldsymbol{\gamma},\mathbf{Y})\) is concave with respect to both \(\boldsymbol{\gamma}\) and \(\mathbf{Y}\), by keeping the other variables fixed, the optimal \(\mathbf{y}_k\) and \(\gamma_k\) are obtained as
    \(\mathbf{y}_k^{\rm opt}=\left(\mathbf{KHH}^{\herm}\mathbf{K}^{\herm}+\frac{\mathbf{KK}^{\herm}}{\rho}+\frac{\mathbf{C_d}}{\rho}\right)^{-1}\mathbf{K}\mathbf{h}_k\) and \(\gamma_k^{\rm opt}={\rm SINDR}_k^{\star}\), respectively.
Now, let us define \(\mathbf{E}\triangleq{\rm blkdiag}[\mathbf{E}_1,\dots,\mathbf{E}_{N_d}]\in \mathbb{B}^{N_{d}N\times N}\), where \(\mathbf{E}_i\) indicates the non-zero locations of each \(i\)th ($i=1,\ldots,N_d$) column of \(\mathbf{Q}^{\herm}\)\cite{DMA_quant_optimization}. By exploiting the relation \({\rm vec}\left(\mathbf{Q}^H\right)=\mathbf{E}\mathbf{q}\), the objective function can be then reformulated with respect to \(\mathbf{q}\). Utilizing the cyclic-shift property of the trace operator and the vectorization properties \cite[1.11.22-1.11.24]{zhang2017matrix}, the optimization problem with respect to \(\mathbf{q}\), similar to \cite[Proposition~3]{DMA_Optimization}, can be formulated as follows:
\vspace{-1mm}
\begin{equation}
\begin{array}{rcl}
\hspace{-1mm} \mathcal{OP}_q\!: & \displaystyle \max_{\mathbf{q}} & \displaystyle \!\!2\Re[\boldsymbol{\xi}^{\herm}\mathbf{q}]\!-\!\mathbf{q}^{\herm}\boldsymbol{\Psi}\mathbf{q}\!-\!\sum_{k=1}^{K}(1\!+\!\gamma_k)\mathbf{y}_k^{\herm}\rho^{-1}\mathbf{C_d}\mathbf{y}_k \\
& \!\!\st & \!\!\mathbf{q}^{\ast}\in \mathcal{L}^{N\times 1},
\end{array}
\end{equation}
where we have used the definitions:
\vspace{-1mm}
\begin{align}
    \boldsymbol{\xi}& \triangleq\bigg(\sum_{k=1}^{K}(1+\gamma_k)(\mathbf{y}_k^{\tran}\mathbf{G}\otimes \mathbf{h}_k^{\herm}\mathbf{A}^{\herm})\mathbf{E}\bigg)^{\herm}, \\
    \nonumber \boldsymbol{\Psi} & \triangleq\mathbf{E}^{\tran}\bigg(\sum_{k=1}^{K}(1+\gamma_k)(\mathbf{Gy}_k^{\ast}\mathbf{y}^{\tran}\mathbf{G})\\
    & \phantom{=} \ \otimes(\mathbf{AHH}^{\herm}\mathbf{A}^{\herm}+\rho^{-1}\mathbf{AA}^{\herm})\bigg) \mathbf{E},
\end{align}
and omitted the first terms, which are invariant to \(\mathbf{q}\). It can be observed that \(\mathbf{G}\) and \(\mathbf{C}_d\) are non-linearly dependent on \(\mathbf{q}\) and, thus, it is hard to deal with them analytically. Next, we approximate them in the regime of large number of users to decouple their dependency on \(\mathbf{q}\).

\begin{figure*}[t]
    \includegraphics[width=\textwidth]{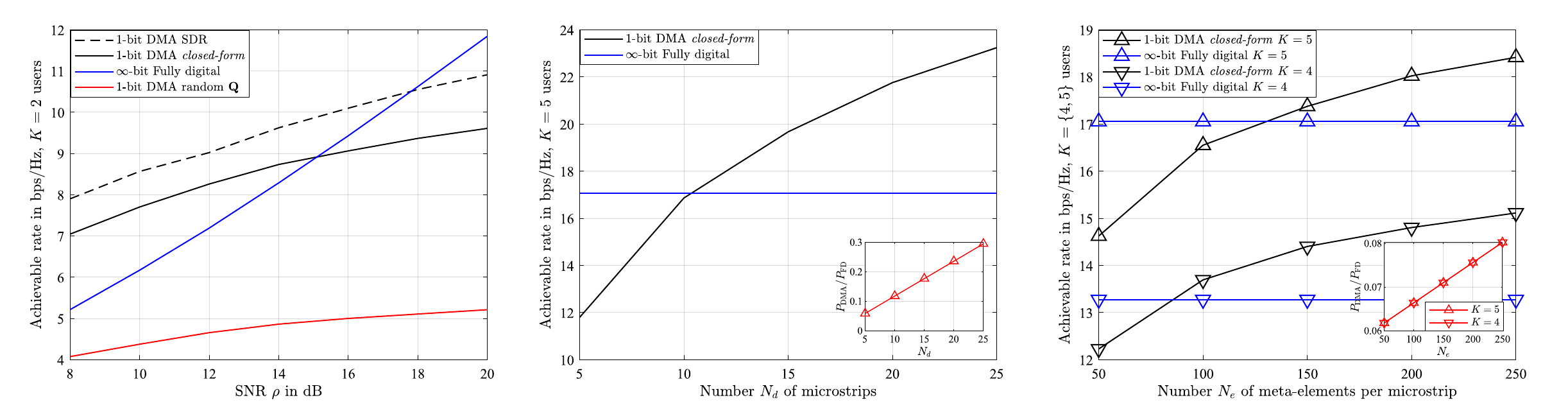}
    \caption{(a) Rate vs \(\rho\) for \(K=2\) users and \(N_{\rm RF}=K\), with \(N_d=5\) and \(N_e=10\) for the 1-bit RX DMA. (b) Rate vs \(N_d\) for \(K=5\) users, \(N_{\rm RF}=K\), \(N_e=20\), and \(\rho=10\) dB. (c) Rate vs \(N_e\) for \(K=\{4,5\}\) users, \(N_{\rm RF}=N_d=K\), and \(\rho=10\) dB. }
    \label{simulation_fig}
    \vspace{-4mm}
\end{figure*}

\vspace{-1mm}
\subsubsection{Approximation of \(\mathbf{G}\) and \(\mathbf{C_d}\) for Large $K$}
\vspace{-1mm}

First, to deal with \(\mathbf{G}\), we need to investigate the structure of the diagonal of \(\mathbf{C_z}\). Hence, we commence by deriving the asymptotic values of the entries of \(\frac{\mathbf{HH}^{\herm}}{K}\).
\begin{prop}\label{Proposition_1}
    For \(K \to \infty\), we have that \(\frac{1}{K} \mathbf{[HH}^{\herm}]_{n,n}\to 1\) and \(\left|\frac{[\mathbf{HH}^{\herm}]_{n,n}}{[\mathbf{HH}^{\herm}]_{i,j}}\right|\to \infty\) \(\forall\, i\neq j\) and \(\forall n,i,j=1,\dots,N\).
\end{prop}
\begin{proof}
    Each diagonal entry of \(\frac{\mathbf{HH}^{\herm}}{K}\) is equal to \(\sum_{k=1}^{K}\frac{|u_{k}|^2}{K}\), where each \(u_k\) is a standard complex normal RV. For \(K \to \infty\) (central limit theorem), we have that \(\sum_{k=1}^{K}\frac{|u_k|^2}{K}\to \Exp[|u_k|^2]=1\). Regarding the off-diagonal entries of \(\frac{\mathbf{HH}^{\herm}}{K}\), we first set \( \varv_k=a_k+\jmath b_k\) and \(v_k=c_k+\jmath d_k\), where \(a_k\), \(b_k\), \(c_k\), and \(d_k\) are zero-mean, $\frac{1}{2}$-variance normal RVs, yielding \(\sum_{k=1}^{K}\frac{\varv_k v_k^{\ast}}{K}\to \Exp[\varv_k v_k^{\ast}]=\Exp[(a_k+\jmath b_k)c_k+(b_k-\jmath a_k)d_k]=0\). Finally, it follows that \(\left|\frac{\sum_{k=1}^{K}\frac{|u_k|^2}{K}}{\sum_{k=1}^{K}\frac{\varv_k v_k^{\ast}}{K}}\right|=\left|\frac{\sum_{k=1}^{K}|u_k|^2}{\sum_{k=1}^{K}u_k v_k^{\ast}}\right|\to \infty\), which concludes the proof.
\end{proof}

The \((n,n)\)th (\(n=1,\dots,N_d\)) entry of \(\mathbf{QAH}\mathbf{H}^{\herm}\mathbf{A}^{\herm}\mathbf{Q}^{\herm}\) can be obtained as follows:
\vspace{-1mm}
\begin{align}\label{eq:Diagonal_entry}
    \nonumber &[\mathbf{QAH}\mathbf{H}^{\herm}\mathbf{A}^{\herm}\mathbf{Q}^{\herm}]_{n,n}=K\sum_{i=1}^{N_e}\alpha_{n, i,i}|q_{n,i}|^2\\
    &+2K\Re\bigg[\sum_{i=1}^{N_e}\sum_{j=i+1}^{N_e}\alpha_{n,i,j}q_{n,i}q_{n,j}^{\ast}e^{\jmath \beta (\ell_{n,i}-\ell_{n,j})}\bigg],
\end{align}
where \(\alpha_{n,i,j}\) corresponds to \(\frac{1}{K}[\mathbf{HH}^{\herm}]_{i+(n-1)N_e,j+(n-1)N_e}\). At this stage, we approximate \(K\sum_{i=1}^{N_e}\alpha_{n, i,i}|q_{n,i}|^2\approx K|\mathbf{q_n}|^2 \), which, under the independence assumption between \(|q_{n,i}|^2\) and \(a_{n,i,i}\) and following Prop.~\ref{Proposition_1}, has a relative error analogous to \(\big|\frac{1}{\sqrt{K}}\big|\). Moreover, \(a_{n,i,i}\gg |a_{n,i,j}|\) \(\forall i\neq j\) and \(a_{n,i,j}\) are zero-mean complex normal RVs. Therefore, the second term is negligible compared to \(K|\mathbf{q}_n|^2\) and \eqref{eq:Diagonal_entry} can be expressed via the following expression:
\vspace{-1mm}
\begin{equation}\label{eq:Aprroximation_1}
    [\mathbf{QAH}\mathbf{H}^{\herm}\mathbf{A}^{\herm}\mathbf{Q}^{\herm}]_{n,n}\approx K |\mathbf{q}_n|^2.
\end{equation}

To ultimately decouple \(\mathbf{G}\) from \(\mathbf{q}\), we need to derive an accurate approximation of \(|\mathbf{q}_1|^2,\dots,|\mathbf{q}_{N_d}|^2\). By exploiting expression \eqref{eq:Lorentzian_constrained}, it follows that:
\vspace{-1mm}
\begin{equation}\label{eq:Approximation_2}
    |\mathbf{q}_n|^2=\frac{1}{2} \bigg(N_e-\sum_{i=1}^{N_e}\sin(\phi_{n,i})\bigg)\approx \frac{1}{2} N_e.
\end{equation}
Then, treating each \(\phi_{n,i}\) as an independent RV uniformly distributed in \([0,2\pi]\), yields \(\sum_{i=1}^{N_e}\sin(\phi_{n,i}) \to 0\) and, for \(N_e \gg 1\), this approximation has a relative error analogous to \(|\frac{1}{\sqrt{2N_e}}|\). Finally, exploiting \eqref{eq:Aprroximation_1} and \eqref{eq:Approximation_2}, \(\mathbf{G}\) can be expressed as \( \mathbf{G}= \sqrt{\frac{4 \eta}{\pi N_e(\rho K+1)}}\mathbf{I}_{N_d}\). Regarding \(\mathbf{C_d}\), it is demonstrated in \cite{Channel_est_1_bit} that, for large values of $K$ or low SNR values, it holds that \( \mathbf{C_d}=\eta \big(1-\frac{2}{\pi} \big)\mathbf{I}_{N_d}\).

Returning to \(\mathcal{OP}_q\), we now focus on its non-convex constraints. Considering that \(\mathbf{q}=\frac{1}{2}(-\jmath \mathbf{e}_{N}+\mathbf{p})\), with \(\mathbf{e}_N\triangleq[1,\dots,1]^{\tran} \in \mathbb{R}^{N \times 1}\) and \(\mathbf{p}\triangleq[e^{\jmath\phi_1},\dots,e^{\jmath \phi_N}]^{\tran}~\forall \phi_i\in[0,2\pi]\), we reformulate the problem similar to \cite{DMA_quant_optimization} as follows:
\vspace{-1mm}
\begin{equation}
\begin{array}{rcl}
    \mathcal{OP}_P: & \displaystyle \max_{\mathbf{P}} & \tr\{\mathbf{MP}\} \\
    & \st & \Diag(\mathbf{P})=\mathbf{I}_{N+1},\,\mathbf{P}\geq\mathbf{0},
\end{array}
\end{equation}
where $\mathbf{P}$, $\mathbf{M} \in \mathbb{C}^{(N+1)\times (N+1)}$ are given by:
\vspace{-1mm}
\begin{equation}
     \mathbf{P}=\begin{bmatrix}
   \mathbf{p} \\
   1 \\
\end{bmatrix}[\mathbf{p}^{\herm}\,1],\; \mathbf{M}=\frac{1}{2}\begin{bmatrix}
    -\mathbf{\Psi} & 2\boldsymbol{\xi}+\jmath \mathbf{\Psi}\mathbf{e}_N \\
    \left(2\boldsymbol{\xi}+\jmath \mathbf{\Psi}\mathbf{e}_N\right)^{\herm} & 0 \\
\end{bmatrix}.
\end{equation}
This problem is a semidefinite relaxation of \(\mathcal{OP}_q\), where we have replaced the rank-one constraint with a positive semidefinite one. Thus, it is a convex problem that can be solved with well-known tools. To obtain its solution, we compute the principal eigenvector of \(\mathbf{P}\), denoted as \(\mathbf{f}^{P}\), and set \(\mathbf{p}^{\rm opt}=e^{\jmath \angle \mathbf{f}^{P}_{1:N}}\). In addition, we propose a \textit{low complexity closed-form approximate solution}, as follows. Considering that the principal eigenvector of \(\mathbf{M}\), denoted as \(\mathbf{f}^{M}\), maximizes the expression \(\mathbf{x}^{\herm}\mathbf{M}\mathbf{x}=\tr\{\mathbf{M}\mathbf{x}\mathbf{x}^{\herm}\}\), we directly set \(\hat{\mathbf{p}}^{\rm opt}=e^{\jmath \angle \mathbf{f}^{M}_{1:N}}\).
%
Finally, we iteratively update \(\boldsymbol{\gamma}\), \(\mathbf{Y}\), and \(\mathbf{p}\) until convergence or until a maximum number of iterations is reached.

\vspace{-1mm}
\section{Numerical Results and Discussion}
\vspace{-2mm}

In this section, we numerically investigate the performance of the proposed 1-bit RX DMA framework for multi-user uplink communications and compare it with a Fully Digital (FD), infinite-resolution MIMO array (referred to in the following simply as ``fully digital'') with $N_{RF}=K$ RX RF chains, each attached to a distinct antenna element. In order to compare the power consumption between the two array architectures, we adopt the following power consumption models: \(P_{\rm DMA}=NP_{\rm el}+N_d(P_{\rm RF}+2 P_{\rm ADC}^{1\text{-}b})\) and \(P_{\rm FD}=N_{\rm RF}(P_{\rm RF}+2 P_{\rm ADC})\), where $b$ is the number of resolution bits and \(P_{\rm ADC}=f_s\, F_{\rm OM}\, 2^b\), with \(\,P_{\rm RF}=60\)~mW, \(f_s=1\)~GHz, \(F_{\rm OM}=500\)~fJ \cite{ADC_FOM_Power_consumpt}, and \(P_{\rm el}=0.1\)~mW \cite{active_ris_larsson}. We consider \(b=10\) bits for the power consumption of the conventional FD architecture.

Figure~\ref{simulation_fig}(a) plots the achievable rate versus \(\rho\) showcasing that the proposed 1-bit RX DMA outperforms the FD architecture up to the SNR value \(\rho \simeq 18\,\)~dB. We also compare the performance of the proposed low complexity closed-form approximate solution and that of random \(\mathbf{Q}\) configurations, demonstrating the effectiveness of our design. It is shown that the approximate solution achieves satisfactory results with significantly lower complexity; in this regard, an optimistic complexity for an interior point method is \(\mathcal{O}\left(N^{6}\right)\) \cite{interior_point_complexity} (\(N^2\) variables), while the low complexity closed-form approximate solution requires \(\mathcal{O}(N^2)\) (power iteration). As for the energy efficiency, considering the adopted models and the simulation parameters in Fig.~\ref{simulation_fig}(a), the 1-bit RX DMA consumes nearly \(7\) times less power. Figure~\ref{simulation_fig}(b) illustrates the achievable rate versus the number of $20$-element microstrips. It is shown that, by employing \(N_d\geq 2N_{\rm RF}\) microstrips, a superior performance is achieved using only $12$--$30\%$ of \(P_{\rm FD}\). This result indicates that, by increasing $N_d$, we can compensate for the 1-bit quantization. In fact, both the rate and the energy efficiency are significantly improved. Lastly, Fig.~\ref{simulation_fig}(c) depicts the rate versus the number of elements per microstrip $N_e$ while keeping \(N_d\) fixed. Again, when increasing $N_e$ and \(N_d=N_{\rm RF}\), the 1-bit RX DMA outperforms the FD architecture. In this regime, only $7$--$8\%$ of \(P_{\rm FD}\) is needed for the proposed scheme to achieve an equal or better sum-rate performance.

\vspace{-3mm}
\section{Conclusion}
\vspace{-2mm}

In this paper, we introduced RX DMAs with 1-bit ADCs for multi-user uplink communications. We presented a closed-form expression for the achievable sum rate, which was used to obtain the optimal analog and digital combiners. It was showcased that, by increasing the number of metamaterials per microstrip, the performance loss due to the 1-bit quantization can be compensated, while achieving a power consumption lower than with conventional MIMO antenna arrays.


\newpage
\bibliographystyle{IEEEbib}
\bibliography{refs}

\begin{thebibliography}{10}

\bibitem{CE_overview_2022}
M.~Jian et~al.,
\newblock ``{Reconfigurable intelligent surfaces for wireless communications:
  Overview of hardware designs, channel models, and estimation techniques},''
\newblock {\em Intel. Converged Netw.}, vol. 3, no. 1, pp. 1--32, 2022.

\bibitem{HMIMO_survey_all}
T.~Gong et~al.,
\newblock ``Holographic {MIMO} communications: {T}heoretical foundations,
  enabling technologies, and future directions,''
\newblock {\em IEEE Commun. Surveys \& Tuts. (to appear)}, 2024.

\bibitem{marconi_george}
C.~Huang et~al.,
\newblock ``Reconfigurable intelligent surfaces for energy efficiency in
  wireless communication,''
\newblock {\em IEEE Trans. Wireless Commun.}, vol. 18, no. 8, pp. 4157--4170,
  2019.

\bibitem{Strinati2021Reconfigurable}
E.~Calvanese~Strinati et~al.,
\newblock ``{Reconfigurable, intelligent, and sustainable wireless environments
  for 6G smart connectivity},''
\newblock {\em IEEE Commun. Mag.}, vol. 59, no. 10, pp. 99--105, 2021.

\bibitem{Space_shift_keying_RIS}
Q.~Li et~al.,
\newblock ``Space shift keying with reconfigurable intelligent surfaces:
  {P}hase configuration designs and performance analysis,''
\newblock {\em IEEE Open J. Commun. Society}, vol. 2, pp. 322--333, 2021.

\bibitem{RIS_pervasive}
G.~C. Alexandropoulos et~al.,
\newblock ``Pervasive machine learning for smart radio environments enabled by
  reconfigurable intelligent surfaces,''
\newblock {\em Proc. IEEE}, vol. 110, no. 9, pp. 1494--1525, Sep. 2022.

\bibitem{PhysFad}
R.~Faqiri et~al.,
\newblock ``{PhysFad}: {P}hysics-based end-to-end channel modeling of
  {RIS}-parametrized environments with adjustable fading,''
\newblock {\em IEEE Trans. Wireless Commun.}, vol. 22, no. 1, pp. 580--595,
  2023.

\bibitem{HMIMO_George}
C.~Huang et~al.,
\newblock ``Holographic {MIMO} surfaces for {6G} wireless networks:
  Opportunities, challenges, and trends,''
\newblock {\em IEEE Wireless Commun.}, vol. 27, no. 5, pp. 118--125, 2020.

\bibitem{DMA_origin}
N.~Shlezinger et~al.,
\newblock ``Dynamic metasurface antennas for uplink massive {MIMO} systems,''
\newblock {\em IEEE Trans. Commun.}, vol. 67, no. 10, pp. 6829--6843, 2019.

\bibitem{gavras2023duplex}
I.~Gavras et~al.,
\newblock ``Full duplex holographic {MIMO} for near-field integrated sensing
  and communications,''
\newblock in {\em Proc. EUSIPCO}, Helsinki, Finland, 2023.

\bibitem{DMA_sensing}
G.~Lan et~al.,
\newblock ``Wireless sensing using dynamic metasurface antennas: Challenges and
  opportunities,''
\newblock {\em IEEE Commun. Mag.}, vol. 58, no. 6, pp. 66--71, 2020.

\bibitem{DMA_quant_optimization}
H.~{Wang} et~al.,
\newblock ``Dynamic metasurface antennas for {MIMO-OFDM} receivers with
  bit-limited {ADCs},''
\newblock {\em IEEE Trans. Commun.}, vol. 69, no. 4, pp. 2643--2659, 2021.

\bibitem{gavras2024dma}
I.~Gavras et~al.,
\newblock ``Near-field localization with 1-bit quantized hybrid {A/D}
  reception,''
\newblock in {\em Proc. IEEE ICASSP}, Seoul, South Korea, 2024.

\bibitem{throughput_analysis_low_bit}
S.~Jacobsson et~al.,
\newblock ``Throughput analysis of massive {MIMO} uplink with low-resolution
  {ADC}s,''
\newblock {\em IEEE Trans. Wireless Commun.}, vol. 16, no. 6, pp. 4038--4051,
  2017.

\bibitem{Channel_est_1_bit}
Y.~Li et~al.,
\newblock ``Channel estimation and performance analysis of one-bit massive
  {MIMO} systems,''
\newblock {\em IEEE Trans. Signal Process.}, vol. 65, no. 15, pp. 4075--4089,
  2017.

\bibitem{Atz22}
I.~{Atzeni} and A.~{Tölli},
\newblock ``Channel estimation and data detection analysis of massive {MIMO}
  with 1-bit {ADCs},''
\newblock {\em IEEE Trans. Wireless Commun.}, vol. 21, no. 6, pp. 3850--3867,
  2022.

\bibitem{THz_1_bit}
R.~Nikbakht and A.~Lozano,
\newblock ``Terahertz transmit beamforming with 1-bit {DAC}s and {ADC}s,''
\newblock in {\em Proc. EUSIPCO}, Dublin, Ireland, 2021.

\bibitem{ADC_FOM_Power_consumpt}
J.~Mo et~al.,
\newblock ``Hybrid architectures with few-bit {ADC} receivers: Achievable rates
  and energy-rate tradeoffs,''
\newblock {\em IEEE Trans. Wireless Commun.}, vol. 16, no. 4, pp. 2274--2287,
  2017.

\bibitem{Atz21}
I.~{Atzeni} et~al.,
\newblock ``Low-resolution massive {MIMO} under hardware power consumption
  constraints,''
\newblock in {\em Proc. Asilomar Conf. Signals, Syst., and Comput.}, Pacific
  Grove, CA, USA, 2021.

\bibitem{Xu_DMA_2022}
J.~Xu et~al.,
\newblock ``Near-field wideband extremely large-scale {MIMO} transmission with
  holographic metasurface antennas,''
\newblock {\em arXiv preprint arXiv:2205.02533}, 2022.

\bibitem{Bussgang_Theory}
O.~D. Demir and E.~Bj\"{o}rnson,
\newblock ``The {B}ussgang decomposition of nonlinear systems: Basic theory and
  {MIMO} extensions [lecture notes],''
\newblock {\em IEEE Signal Process. Mag.}, vol. 38, no. 1, pp. 131--136, 2021.

\bibitem{heath2018foundations}
R.~W. Heath and A.~Lozano,
\newblock {\em Foundations of MIMO Communication},
\newblock Cambridge University Press, 2018.

\bibitem{DMA_Optimization}
H.~Wang,
\newblock ``Joint transmitter and receiver design for uplink {MU-MIMO} systems
  with dynamic metasurface antennas,''
\newblock in {\em Proc. IEEE VTC-Spring}, Helsinki, Finland, Jun. 2022.

\bibitem{zhang2017matrix}
X.-D. Zhang,
\newblock {\em Matrix Analysis and Applications},
\newblock Cambridge University Press, 2017.

\bibitem{active_ris_larsson}
R.~Long et~al.,
\newblock ``Active reconfigurable intelligent surface-aided wireless
  communications,''
\newblock {\em IEEE Trans. Wireless Commun.}, vol. 20, no. 8, pp. 4962--4975,
  2021.

\bibitem{interior_point_complexity}
L.~Vandenberghe et~al.,
\newblock ``Interior-point algorithms for semidefinite programming problems
  derived from the {KYP} lemma,''
\newblock in {\em Positive Polynomials in Control}, pp. 195--238. Springer,
  2005.

\end{thebibliography}

\end{document}